\documentstyle[12pt]{article}
\newcommand{\AmS}{{\protect\the\textfont2
\renewcommand{\thesection}{\Roman{section}}
  A\kern-.1667em\lower.5ex\hbox{M}\kern-.125emS}}
 
\begin{document}
%\baselineskip 24 pt
%\rightline {\today}
\rightline {CERN-TH/96-195}
%\rightline {hep-lat/9607???}
\vskip 2. truecm
\centerline{\bf Strongly Coupled QED$^*$}
\vskip 2 truecm
\centerline { Vicente ~Azcoiti}
\vskip 1 truecm
\centerline {\it CERN Theory Division,}
\centerline {\it CH-1211, Geneve 23, Switzerland}
\vskip 3 truecm
\centerline {ABSTRACT}
A short review of some of the most relevant contributions to non-perturbative 
QED is done. Since a Gaussian behaviour of QED \`a la $\lambda\phi^4$ 
has been ruled out by the numerical data, I analyse the other two most 
reliable scenarios, i.e. triviality \`a la Nambu-Jona Lasinio and 
non-Gaussian critical behaviour. I give a suggestive theoretical argument 
against a Gaussian behaviour of QED \`a la Nambu-Jona Lasinio, and show 
how the numerical data for the susceptibility at the critical point of 
QED support this result.
\vskip 6truecm
\noindent
{$^*$ \it Invited Plenary Talk given at the Lattice 96 Symposium, 
St. Louis, Missouri, June 1996.}
\vfill
\begin{flushleft}
CERN-TH/96-195\\
July 1996
\end{flushleft}
\eject

\par
\leftline{\bf 1. Introduction}
\par
Quantum Electro-Dynamics (QED) is the most standard and popular example 
of a non-asymptotically free gauge theory. Due to the peculiar form of its 
$\beta$ renormalization group function, as obtained in perturbation 
theory, this model suffers from the Landau pole problem, which means that 
the renormalized coupling increases with the energy and diverges at some 
finite energy scale. The only way to construct a consistent theory 
in the infinite cut-off limit is to put the renormalized coupling equal 
to zero and the model is therefore trivial. In spite of that, QED is 
also the most successful quantum field theory from a phenomenological 
point of view. In fact, at energy scales much lower than the cut-off 
scale, QED describes perfectly the electromagnetic interactions. There 
is no doubt that QED incorporates very well the main features of 
electromagnetic interactions between electrons and photons in Nature.

There are several possible issues to the triviality problem of perturbative 
QED:

i. Non-perturbative contributions to the $\beta$ function, which would be 
negligible at small energies, become relevant at large energy scales and 
change the structure of the $\beta$ function. An ultraviolet stable or 
non-Gaussian fixed point could appear in this way as driving 
to a non-trivial 
interacting theory in the infinite cut-off limit. This issue includes also 
the possibility that a summation of all the perturbative series be able 
to change the structure of the $\beta$ function. Therefore 
``non-perturbative 
contributions'' should be understood in a wide sense.

ii. At large energy scales, and in the context of Unification Models for 
gauge interactions, other physical interactions between particles become 
relevant; these new interactions are responsible for the necessary 
changes in the $\beta$ function in order to get a non-trivial model.

iii. The real world is only described by effective theories, 
which are interacting 
at small energy scales compared with the cut-off scale but are trivial if 
the cut-off is sent to infinity.

It could be that, after all, the perturbative Landau pole problem remains, 
even taking into account all ``non-perturbative'' contributions to the 
$\beta$ function. However, and independently of this, it 
is very interesting and stimulating, 
at least from a theoretical point of view, 
to understand if 
non-asymptotically free gauge theories can be consistently constructed 
against the standard prejudice, and to look for the possible existence 
of non Gaussian fixed points in QED. This is however a hard 
business since, contrary to what happens in asymptotically free gauge 
theories, perturbative solutions of the renormalization group 
equations cannot be used as a guide in these investigations.

Another source of interest in non-perturbative QED comes from the 
composite models for the Higgs sector of the Standard Model. In fact 
it has been argued that a technigauge model able to strongly couple 
fermions and to produce massive fermion-antifermion bound states at 
large energy and with large anomalous dimensions (not asymptotically 
free!) could be on the basis of the dynamical mass generation 
mechanism.

These are the main reasons that motivated people $12$ years ago [1] 
to start investigations on non-perturbative QED and to search for 
the existence of a non-Gaussian fixed point in this model. An important 
amount of work in this direction has been done in these $12$ years. 
References [1-11] are some of the most relevant contributions to the 
field.

The existence of a continuous chiral transition at finite inverse gauge 
coupling $\beta$, separating a strongly coupled phase where chiral 
symmetry is broken from a weak coupling phase where the symmetry is 
realized, was found in [1] for the non-compact model in the quenched 
approximation. Another very interesting and important contribution 
to the field was the work by Leung, Love and Bardeen [3]: using 
the continuum formulation and solving the Schwinger-Dyson equations 
for the fermion self-energy in the quenched ladder approximation, 
they found 
large anomalous dimensions near the critical point for the composite 
operator $\bar\psi\psi$; this implies a scale dimension 
of 2 for $\bar\psi\psi$ and strongly suggests that four-fermion 
interactions become renormalizable outside perturbation theory. 
We should therefore include four-fermion interactions in a full 
analysis of non perturbative QED.

Reference [4] was the first numerical simulation of the compact model 
with dynamical fermions whereas in [5] it was shown that the compact 
model undergoes a first order chiral phase transition and therefore 
a continuum limit cannot be defined for this model. This point deserves 
further investigations since, as recently found [12], a continuous chiral 
transition appears in the compact model after the inclusion of four 
fermion interactions.

Reference [6] contains the results of the first numerical simulation of 
non-compact QED with dynamical fermions. In [7] it was shown that the 
chiral condensate follows a power law behaviour near the critical point, 
as opposed to the essential singularity behaviour or Miransky scaling [2]. 

The DESY-J\"ulich group claimed in [8] that they were able to fit the data 
for the chiral condensate with a logarithmically improved scalar mean 
field equation of state. This was a five-parameter fit, one of them being 
the critical coupling. From the results of these fits they concluded that 
QED has a trivial continuum limit. Triviality would be described by 
a Gaussian fixed point, the critical behaviour around it being similar 
to the one of the $\lambda\phi^4$ model. In reference [9], two 
completely independent and uncorrelated calculations 
were done using different 
numerical approaches and different operators. The extracted 
critical couplings, which were in perfect agreement, did not support 
however the logarithmically improved scalar mean field fit of the 
DESY-J\"ulich group. The data of [9] were compatible, as discussed 
below, with a Gaussian behaviour of QED \`a la Nambu-Jona Lasinio as 
well as with a pure power law equation of state. 

Actually, the high degree of disagreement between the 
numerical estimations of the critical coupling in [8] and [9] 
is surprising, a 
disagreement that in our more recent calculations in the $14^4$ 
lattice is of about 100 standard deviations [13]. Since the numerical 
data for the chiral condensate in both calculations were in good 
agreement, the only source of disagreement must be in the 
analysis of these data, i.e. in the use of different fitting 
equations in [8] and [9]. The DESY-J\"ulich group has recently 
published new results on the logarithmically improved scalar mean 
field fits of QED with higher statistics [14]. In this paper the 
$\frac{\chi^2}{d.o.f.}$ of the fits is also shown. 
The best reported value is 
$\frac{\chi^2}{d.o.f.}= 7.63$ for around 70 degrees of freedom. These 
numbers give an extremely low confidence level for these fits and 
rule out a Gaussian behaviour of QED \`a la $\lambda\phi^4$. Indeed, 
as suggested by Horowitz a few years ago [10], if the continuum limit 
of QED is trivial, triviality would manifest itself as in the Nambu-Jona 
Lasinio model rather than as in the $\lambda\phi^4$ model.

It is simple to understand this statement if we take it into account that 
QED in the infinite gauge coupling limit and Nambu-Jona Lasinio in the 
infinite four-fermion coupling limit are equivalent. Both models have 
composite scalars and pseudo-scalars in the spectrum and, as suggested 
by Koci\'c and Kogut, and corroborated in a numerical simulation of 
the Nambu-Jona Lasinio model with discrete $Z_2$ symmetry [10], triviality 
in a theory with composite scalars seems not to manifest itself 
in the same way as 
in a theory with fundamental scalars. This result can be shown 
for the Nambu-Jona Lasinio model in a very simple and elegant way by 
means of the large $N$ expansion.

Reference [11] contains the results of an analysis of the gauged 
Nambu-Jona Lasinio model, i.e. QED plus four-fermion interactions 
with continuous chiral symmetry. A mean field approximation for the 
fermion field was used in [11], whereas fluctuations of the gauge field 
were taken into account outside of approximations. The finding of a 
continuous phase transition line of critical points 
in the gauge-coupling--four-fermion-coupling plane, with critical indices 
taking non-mean field values, at least 
near the critical point of QED [11], 
was very suggestive. Notwithstanding that we would like to understand 
the nature of the continuum limit of strongly coupled QED outside 
of approximations.

I will develop in the next section a suggestive argument against a 
Gaussian behaviour of QED \`a la Nambu-Jona Lasinio. In section 3 
I will discuss some recent results of the Frascati-Zaragoza collaboration 
on the chiral equation of state of QED at criticality, and the last 
section will be devoted to the conclusions. Even if I am not going to 
discuss it here, I would like to state that there are other interesting 
approaches for the pure gauge compact model based on a suppression 
of monopole contributions [15, 16], the results being 
appealing enough to deserve further investigations.

\vskip 1truecm
\leftline{\bf 2. The Effective Action}
\par
As discussed in the previous section, a Gaussian behaviour of QED as in 
a logarithmically improved scalar mean field theory has been ruled out 
[9, 13], also by the very low confidence level of the fits reported in [14]. 
This notwithstanding, a Gaussian behaviour of QED \`a la Nambu-Jona Lasinio 
seems to be compatible with the data and it is a difficult task 
to distinguish this behaviour from 
a non-Gaussian behaviour. Indeed in both cases the 
effective $\delta$ exponent which measures the response of the system to 
a external symmetry breaking field at the critical point is less than 3. 
The effective magnetic exponent $\beta_m$, which controls the behaviour 
of the chiral condensate near the critical point, is larger than its 
mean field value of $0.5$ in both cases; the data for the susceptibility 
exponent $\gamma$ point clearly to the mean field value $\gamma=1$ [17]. 
To differentiate between logarithmic violations to mean field scaling 
and a pure power law scaling is actually very difficult (see next 
section).
This is the reason why, before discussing the numerical results, I would 
like to give an argument based on theoretical grounds, strongly suggesting 
that QED and Nambu-Jona Lasinio are in different universality classes. 
This argument is based on the definition of an effective action for the 
Nambu-Jona Lasinio model and on its comparison with the fermion effective 
action of QED at fixed energy density.

As we have shown in the past [9, 17], the critical behaviour of QED is 
controlled by a singularity of the fermion effective action, which holds 
at a critical value of the gauge energy density $E$ around 1. I will now 
show how we can define an effective action for the Nambu-Jona Lasinio 
model, the strong four-fermion coupling expansion of which has as first 
contribution the QED effective action. Before that, let me review the 
main features of the construction of the fermion effective action in 
QED.

The lattice action of non-compact QED with staggered fermions is

$$S=\frac{\beta}{2}\sum_{n,\mu<\nu}F_{\mu\nu}^2(n)\:+\:
\frac{1}{2}\sum_{n,\mu}\eta_{\mu}(n)(\bar\psi_nU_{\mu}(n)\psi_{n+\mu}
-\bar\psi_{n+\mu}U_{\mu}^+(n)\psi_{n}) + 
m\sum_{n}\bar\psi_{n}\psi_{n}.
\eqno(1)$$

\noindent
where $\beta=1/g^2$, $F_{\mu\nu}$ is the naive regularization of the 
continuum e.m. tensor, $\eta_{\mu}(n)$ are the Kogut-Susskind phases 
and $U_{\mu}(n)$ are the compact link connections. In the chiral limit 
this action is invariant under the continuous chiral $U(1)$ 
transformations

$$\psi_n\;\;\rightarrow\;\;\psi_n\: e^{i\alpha (-1)^{n_1+\ldots+n_d}}
\;\;\;\;\;\;\;
\;\;\bar\psi_n\;\;\rightarrow\;\;\bar\psi_n\: e^{i\alpha (-1)^{n_1+\ldots
+n_d}}.
\eqno(2)$$

Action (1) can be written in a compact form as 

$$S=\frac{\beta}{2}\sum_{n,\mu<\nu}F_{\mu\nu}^2(n)\:+\:\bar\psi\Delta
(m,U_{\mu}(n))\psi
\eqno(3)$$

\noindent
with the following form of the Dirac operator

$$\Delta = mI + i\Lambda(U_{\mu}(n))
\eqno(4)$$

\noindent
$\Lambda(U_{\mu}(n))$ being a Hermitian matrix. After integrating out the 
fermion degrees of freedom the partition function is

$$
Z =\int [dA_{\mu}(n)] e^{-\frac{\beta}{2}\sum_{n,\mu<\nu}F_{\mu\nu}^2(n)} 
\det \Delta(m, U_{\mu}(n))
\eqno(5)$$

In order to construct the fermion effective action at fixed gauge energy 
density we introduce in the previous expression a trivial identity in the 
following way [18]: 

$$Z =\int [dA_{\mu}(n)] dE \delta\left(\frac{1}{12V}\sum F_{\mu\nu}^2(n)
-E\right) 
e^{-\frac{\beta}{2}\sum_{n,\mu<\nu}F_{\mu\nu}^2(n)} det\Delta
\eqno(6)$$

\noindent
where $V$ in (6) is the lattice volume. If we now introduce the density of 
states $N(E)$ at fixed gauge energy density as 

$$
N(E) =\int [dA_{\mu}(n)]\delta\left(\frac{1}{12V}\sum F_{\mu\nu}^2(n)
-E\right),  
\eqno(7)$$

\noindent
the partition function (6) can be written as

$$
Z=\int dE N(E) e^{-V6\beta E} {\langle\,{\det \Delta}\rangle}_{E}, 
\eqno(8)$$

\noindent
with the following definition for the mean value of the fermion determinant 
at fixed $E$: 

$$
{{\langle\,\det \Delta\rangle}_{E}} = 
\frac{\int [dA_{\mu}(n)] \det\Delta \delta(\frac{1}{12V}\sum F_{\mu\nu}^2(n)
-E) }{\int [dA_{\mu}(n)]\delta(\frac{1}{12V}\sum F_{\mu\nu}^2(n)-E) }. 
\eqno(9)$$

Contrary to what happens in the compact case, the partition function (6), 
(8) is divergent due to gauge invariance and to the use of non-compact 
variables. These divergences, which are avoidable by fixing the gauge, 
can also be regularized, and they cancel when taking vacuum expectation 
values.

Since the pure gauge action is a quadratic form of the gauge fields, the 
density of states $N(E)$ can be analytically computed [9]. At large 
lattice volumes it behaves like

$$
N(E) = C E^{V\frac{3}{2}}, 
\eqno(10)$$

\noindent
where $C$ is a divergent constant that contains the divergences due to 
the non-compact integration, and which cancel when computing mean values.

Now taking into account the fact that the fermion 
determinant is positive-definite, 
we can introduce the normalized effective action $S_{eff}^{QED}(E,m)$ as 

$$
{S_{eff}^{QED}(E,m)} = -\frac{1}{V} \log {\langle\,{\det \Delta}\rangle}_{E}, 
\eqno(11)
$$

\noindent
which takes a finite value in the thermodynamical limit. Expressions (8), 
(10) and (11) allow us to write the partition function as a simple 
one-dimensional integral over the gauge energy density: 

$$
Z = \int dE e^{V\frac{3}{2}log E-V6\beta E- V{S_{eff}^{QED}(E,m)}}. 
\eqno(12)
$$

The only unknown function in the previous expression is the effective action 
${S_{eff}^{QED}(E,m)}$. It contains all the information on the critical 
behaviour of this model.

The thermodynamics of this model can be exactly solved in the infinite 
volume limit by the saddle point technique [9]. The mean plaquette energy 
for instance 
will be given by the solution of the saddle point equation, which 
maximizes the integrand in (12).

Let me now consider the chiral limit. In this limit, the model has a 
continuous phase transition at some finite value of the gauge coupling,  
in perfect analogy with magnetic systems. Since first-order phase 
transitions are excluded at small numbers of flavours [19, 20] it is obvious 
that if the effective action ${S_{eff}^{QED}(E)}$ is a regular function 
of $E$ the solutions of the saddle point equations will also be regular.
No singularities in the plaquette energy or specific heat will be found 
at any finite $\beta$. Therefore in order to get a continuous phase 
transition at finite $\beta$ we need a singularity in the effective action 
${S_{eff}^{QED}(E)}$ at some finite value of $E$. This was just what we 
found by numerical computation of ${S_{eff}^{QED}(E)}$ [9]. 

The investigation of the main features of this singularity by numerical 
methods is rather difficult. However, what is important for the argument 
I am going to develop here is that this singularity controls the critical 
behaviour of the model and determines its critical exponents.

The next step in this argument is to define an effective action for the 
Nambu-Jona Lasinio model and to compare it with the QED effective action.
The lattice action for the Nambu-Jona Lasinio model with continuous chiral 
symmetry and in the chiral limit is

$$
{S_{NJL}}=
\frac{1}{2}\sum_{n,\mu}\eta_{\mu}(n)(\bar\psi_n\psi_{n+\mu}
-\bar\psi_{n+\mu}\psi_{n})
\:-\:G\sum_{n,\mu}\bar\psi_n\psi_n\bar\psi_{n+\mu}\psi_{n+\mu} 
\eqno(13)
$$

\noindent
where $G$ is the four-fermion coupling. This action is also invariant 
under the continuous chiral transformations (2).

We can bilinearize the previous action with the help of an auxiliary 
compact $U(1)$ vector field as follows

$$
{S_{NJL}}=
\frac{1}{2}\sum_{n,\mu}\eta_{\mu}(n)(\bar\psi_n\psi_{n+\mu}
-\bar\psi_{n+\mu}\psi_{n})
$$
$$
\:+\:G^{\frac{1}{2}}
\sum_{n,\mu}\eta_{\mu}(n)(\bar\psi_ne^{i\theta_{\mu}^{F}(n)}\psi_{n+\mu}
-\bar\psi_{n+\mu}e^{-i\theta_{\mu}^{F}(n)}\psi_{n}).
\eqno(14)$$

\noindent
$\theta_{\mu}^{F}(n)$ are compact phases taking values in the 
$(-\pi,\pi)$ interval. Action (14) can be written in a compact form as 

$$
{S_{NJL}}= 
\bar\psi\Delta_{NJL}(\theta_{\mu}^{F})\psi 
\eqno(15)$$

\noindent
with the following expression for the Dirac operator: 

$$
{\Delta_{NJL}} = \Delta_{0}\:+\:(4G)^{\frac{1}{2}}
\Delta(\theta_{\mu}^{F}).
\eqno(16)$$

\noindent
$\Delta_{0}$ is the Dirac operator for a free fermion theory, whereas 
$\Delta(\theta_{\mu}^{F})$ is the QED Dirac operator with the gauge field 
$A_\mu$ replaced by the auxiliary vector field $\theta_{\mu}^{F}$.

The partition function is the path integral over the auxiliary vector field 
configurations of the determinant of the Dirac operator

$$
{Z_{NJL}} = 
\int [d\theta^{F}] \det \Delta_{NJL}(\theta_{\mu}^{F}).
\eqno(17)$$

Since the fermion determinant in (17) is a periodic function of the 
compact variables $\theta_{\mu}^{F}$, we can also use non-compact variables 
in the definition of $Z_{NJL}$. The divergences that will appear from 
the use of non-compact degrees of freedom can be regularized; they cancel 
when computing vacuum expectation values, as in non-compact QED.

The next steps are the same as were used before in the definition of the QED 
effective action. First we define the energy density of the auxiliary vector 
field $\theta_{\mu}^{F}$ just as if it were a gauge field with a 
``e.m. tensor'': 

$$
{F_{\mu\nu}^{F}(n)} = 
\theta_{\mu}^{F}(n)\:+\:\theta_{\nu}^{F}(n+\mu)\:-\:\theta_{\mu}^{F}(n+\nu)
\:-\:\theta_{\nu}^{F}(n). 
\eqno(18)$$

Inserting now in the partition function (17) a $\delta$ function 

$$
\delta\left(\frac{1}{12V}\sum_{n,\mu<\nu}F_{\mu\nu}^2(n)-E\right)
$$

\noindent
and an integral over the energy density $E$ of the auxiliary vector field, 
we get

$$
Z_{NJL} = 
\int dE e^{V\frac{3}{2}\log E - V S_{eff}^{NJL}(E,G)}
\eqno(19)$$

\noindent
with

$$
{S_{eff}^{NJL}(E,G)} = 
-\frac{1}{V}\log{\langle\,{\det \Delta_{NJL}}\rangle}_{E}, 
\eqno(20)$$

\noindent
and where the definition of the mean value 
${\langle\,\rangle}_{E}$ over configurations at fixed energy density 
of the vector field $\theta_{\mu}^{F}$ is the same as given in (9).

Again in this case the thermodynamics of the system can be exactly 
solved in the infinite volume limit by the saddle point technique. 
For instance, the mean value of the four-fermion term in the action (13) 
will be given by 

$$
{\langle\,\bar\psi_n\psi_n\bar\psi_{n+\mu}\psi_{n+\mu},\rangle}=
\frac{\partial S_{eff}^{NJL}(E,G)}{\partial G}
\eqno(21)$$

\noindent
evaluated at the solution $E=E_{0}(G)$ of the saddle point equation 
for each fixed value of the four-fermion coupling $G$.

Doing now a cumulant expansion of 
${{\langle\,\det \Delta_{NJL},\rangle}_{E}}$ and inserting in it the expression 
(16) for the Nambu-Jona Lasinio Dirac operator, we get a strong four-fermion 
coupling expansion for the Nambu-Jona Lasinio effective action

$$
{S_{eff}^{NJL}(E,G)} = 
{S_{eff}^{QED}(E)} \:+\:{S_{1}(E,G)}. 
\eqno(23)$$

The first contribution to this expansion ${S_{eff}^{QED}(E)}$ is just the 
effective action of QED, whereas ${S_{1}(E,G)}$ in (23) vanishes in the 
$G=\infty$ limit.

We have now at hand everything we need to finish the argument. As discussed 
before, the QED effective action has a singularity that holds at a 
critical value of the energy $E$ around 1 [9, 17] 
and which controls the critical 
behaviour of the model. Since the QED effective action is the first 
contribution to the Nambu-Jona Lasinio effective action (23) in the strong 
four fermion coupling expansion of it, the QED singularity will also be 
a singularity of the Nambu-Jona Lasinio effective action. Hence, if by running 
the four-fermion coupling $G$ the solution of the saddle point equations 
$E_{0}(G)$ for this model reaches the critical energy of QED, the critical 
behaviour of the Nambu-Jona Lasinio model will be controlled also by the same 
singularity that controls the critical behaviour in QED. Both models will 
be in the same universality class even if the behaviour far from the critical 
point could be very different.

What breaks the previous argument is the assumption that the singularity of 
QED is accessible to the Nambu-Jona Lasinio model. In fact since there is 
no kinetic term for the auxiliary vector field in action (14) and since 
the Nambu-Jona Lasinio effective action is bounded, the solution of the 
saddle point equation is $E_{0}(G)=\infty$ for every value of the 
four-fermion coupling $G$. The singularity of QED is not accessible to the 
Nambu-Jona Lasinio model. The critical behaviour of the last model 
must be controlled by a singularity in the four-fermion coupling $G$ of 
its effective action at $E=\infty$.

The argument I have developed here has a close resemblance with real 
space renormalization group based arguments. Even if it does not 
constitute a rigorous proof, it strongly suggests that both models are 
in different universality classes.

\vskip 1truecm
\leftline{\bf 3. The QED Chiral Equation of State at Criticality}
\par
In this section I will discuss some recent results on the behaviour of the 
chiral order parameter at the critical point of QED [13]. The main 
ingredients of this calculation are the use of the Microcanonical Fermion 
Average approach (MFA) to simulate dynamical fermions [18], and the 
computation of the chiral transverse susceptibility in the Coulomb phase 
and in the chiral limit, without the help of arbitrary mass extrapolations.

In the MFA approach the massless Dirac operator is exactly diagonalized 
with the help of a modified Lanczos algorithm. In contrast to what happens 
with other standard algorithms such as Hybrid Monte Carlo, computations in 
the chiral limit are therefore allowed in this approach. We only need to 
take care that the commutation of the chiral and thermodynamical limits 
can be done for the operator we are interested in. In the broken phase, 
characterized by a non-symmetric vacuum under chiral transformations, 
this commutation will give wrong results if applied to operators 
which are not invariant under chiral transformations. However the symmetry 
of the vacuum under chiral transformations allows in principle such a 
commutation in the Coulomb phase. This result can be rigorously demonstrated, 
for instance for the transverse susceptibility. The only ingredient in this 
demonstration is the fact that the susceptibility in the Coulomb phase and 
in the chiral limit is always finite, which implies that the 
spectral density of eigenvalues of the Dirac operator $\rho(\lambda)$ 
near the origin behaves like $\lambda^p$, 
with $p>1$. Using this result, it is simple 
to show that the transverse susceptibility is a continuous function of 
the bare fermion mass $m$ at $m=0$.

The computation of the chiral susceptibility in the Coulomb phase and in the 
chiral limit allowed us to get very precise measurements of the critical 
couplings [13, 17]. This is a crucial point since, as discussed in [9], the 
numerical determination of critical indices is very sensitive to the value 
of the critical coupling. As I mentioned in the introduction, the numerical 
estimations of the critical coupling by the Illinois group and the 
Frascati-Zaragoza group [9] were in very good agreement even if numerical 
approaches and techniques used in these calculations were very different. 
I would like to mention here that we have four independent determinations 
of the critical coupling based on the analysis of the fermion effective 
action [9], probability distribution function of the chiral order 
parameter [21], Lee-Yang zeros of the partition function and chiral 
susceptibilities [13, 17]. The results of the four calculations 
were in perfect agreement, but the values extracted from the computation 
of the chiral susceptibility are the best from a statistical point of 
view. The reason for that was the observation that chiral susceptibility 
follows an almost perfect linear behaviour as a function of the gauge 
energy density in the Coulomb phase, even relatively far from the 
critical point. The very high confidence level of the linear fits for 
this operator are responsible for the very precise measurements of 
the critical couplings (see Table I).

Using the values for the critical couplings reported in Table I, fits of 
the data for the chiral order parameter at the critical point were done 
with the two most reliable equations of state. As previously stated, a 
Gaussian behaviour of strongly coupled QED \`a la $\lambda\Phi^4$ has 
been ruled out by the numerical data [9, 13, 14]. I also discussed how, 
from theoretical prejudices, we expect that, if QED is trivial, triviality 
would manifest itself as in the Nambu-Jona Lasinio model. Even if I have 
given in the previous section a strong argument against a Gaussian 
behaviour of strongly coupled QED \`a la Nambu-Jona Lasinio, I will show 
here the results of the fits of the chiral order parameter with a 
Nambu-Jona Lasinio equation of state and with a pure power law, the last 
corresponding to a non-Gaussian behaviour of the model.

The Nambu-Jona Lasinio-like equation of state at the critical point is 
parametrized as follows

$$A{\langle\,\bar\psi\psi\,\rangle^3}{\log\left({1\over{
\langle\,\bar\psi\psi\,\rangle}}\right) + 
C\langle\,\bar\psi\psi\,\rangle^3}
= m. 
\eqno(24)$$

\noindent
The pure cubic term in this equation fixes the scale of the logarithmic 
violations to scaling. The power of the logarithm is set equal to 1, as 
follows from the $\frac{1}{N}$ expansion of the Nambu-Jona Lasinio model 
[22].

The pure power law equation of state has the following simple form

$$B{\langle\,\bar\psi\psi\,\rangle^{\delta}}
= m. 
\eqno(25)$$

The fits of Figs. 1 and 2 have been done in the mass range $0.01<m<0.08$ 
and $14^4$ lattice. In both Figs. we plot the left-hand side of the 
corresponding equation of state against the right-hand side. Very high 
quality fits are obtained in the two cases, as follows from the results 
reported in Table II, where I show the $\frac{\chi^2}{d.o.f.}$ 
of the fits in 
$10^4,12^4$ and $14^4$ lattices and for a number of flavours running 
from 1 to 8. The errors in this table take into account the error in 
the determination of the critical coupling (see Table I).

The conclusion of this analysis is that the data for the chiral condensate 
at the critical point are not enough sensitive to distinguish between the 
two possibilities. Both equations of state, with the corresponding fitted 
parameters, are actually very similar in the mass region explored and 
even at much smaller masses.

The next step in complexity of the numerical analysis is to test the 
behaviour of the derivative of the order parameter, i.e. the longitudinal 
susceptibility. We expect this operator, being a higher order derivative 
of the free energy, to be more sensitive to different equations of state.

In the case of a Nambu-Jona Lasinio-like equation of state we get for the 
longitudinal susceptibility $\chi_{L}$ the following equation

$${3m - \langle\,\bar\psi\psi\,\rangle\chi^{-1}} = 
{A {\langle\,\bar\psi\psi\,\rangle^{3}}}
\eqno(26)$$

\noindent
whereas for the pure power law the equation is

$${3m - \langle\,\bar\psi\psi\,\rangle\chi^{-1}} = 
{(3-\delta) B {\langle\,\bar\psi\psi\,\rangle^{\delta}}}. 
\eqno(27)$$

Equations (26) and (27) have two remarkable things. First, the 
logarithmic violations to scaling of equation (24) disappear in the 
corresponding equation for the susceptibility (26). This is very 
welcome from a numerical point of view since it is very difficult 
to distinguish $\delta=3$ 
with logarithmic violations driving to a $\delta_{eff}<3$ from $\delta<3$. 
Second, if we do a log--log plot of 
$\log {(3m - \langle\,\bar\psi\psi\,\rangle\chi^{-1})}$ against 
$\log {\langle\,\bar\psi\psi\,\rangle}$ we expect a linear behaviour in both 
cases (26), (27). The only difference is the slope of the linear fit, 
which should be 3 in the Gaussian case and less than 3 for the 
non-Gaussian one.

Figure 3 contains the results for the log--log plot in the $12^4$ lattice. 
All the plotted points in this Fig. are in the mass range $(0.015--0.035)$. 
The straight line in this Fig. is a linear fit. From the slope of this 
fit we get that the mean field-like equation of state (Nambu-Jona 
Lasinio) is also ruled out since the extracted value for the critical 
index is $\delta= 1.98(16)$.

These results, which are very appealing, unfortunately need to be improved 
in larger lattices before they can be taken 
as a ``proof'' of the existence of 
a non-Gaussian fixed point in strongly coupled QED. We have not 
enough statistics to check the previous results in the $14^4$ lattice. 
This notwithstanding, the result $\delta= 1.98(16)$, which is very similar 
to the quenched value reported in [23], is far enough from the Gaussian 
$\delta=3$ value to consider it as a strong indication of a non-Gaussian 
behaviour.

To finish this section I would like to comment on some features of Fig. 3. 
We have found deviations from the linear behaviour of the function 
plotted in this Fig. in both the left-hand and right-hand corners. 
The right-hand corner corresponds to very small bare 
fermion masses, and we expect deviations from scaling in this 
region due to finite-size effects. This is the region we should explore 
on larger lattices in order to see if the scaling window enlarges to 
smaller fermion masses.

The physical origin for the deviations from scaling in the left-hand corner 
is different. In this case we have large bare fermion masses and it is most 
likely that deviations from scaling come from subleading contributions to 
the dominant scaling in equations (25), (27).

Another surprising feature, which follows from the analysis of the results 
plotted in Fig. 3, is the value $\delta= 1.98(16)$. This value is quite 
different from the value $\delta=2.81$ obtained for the same lattice size 
from the fits of the chiral order parameter with the pure power law 
equation (25). To understand this discrepancy we should take into account 
that the fits of equation (25) were done in a much larger mass interval 
than the one used in Fig. 3. As stated before, the chiral order parameter 
is not very sensitive to the different equations of state. Playing with the 
two free parameters of equation (25), we can of course reproduce the data 
for the chiral condensate in the mass interval $(0.015--0.035)$ and 
$\delta\sim 2$. The susceptibility fits, as expected, are much more sensitive 
to the form of the equation of state, even if the mass interval were these 
fits work well is significantly reduced.

\vskip 1truecm
\leftline{\bf 4. Summary}
\par
I have done in the Introduction of this paper a short review of some of the 
most relevant contributions to the investigation of non-perturbative QED 
[1--11]. The relevance of four-fermion interactions in this kind of 
investigations [3] has been emphasized. Unfortunately, and due to 
technical difficulties, results for the full model with four-fermion 
interactions are available only in the mean field approximation for the 
fermion field [11]. These results are however very exciting, since 
a non-Gaussian behaviour appears in this approximation [11].

A Gaussian behaviour of QED \`a la $\lambda\phi^4$ has been ruled out by the 
results of the Illinois and Frascati-Zaragoza groups [9, 13] as well as by the 
extremely low confidence level of the logarithmically improved scalar mean 
field fits of the DESY-J\"ulich group [14]. These results, beside theoretical 
prejudices based on the equivalence between QED and Nambu-Jona Lasinio 
models in the strong coupling limit, tell us that the most reliable 
realization of triviality in QED, assuming it is trivial, is \`a la 
Nambu-Jona Lasinio.

I have given in section 2 a suggestive theoretical argument against a Gaussian 
behaviour of QED \`a la Nambu-Jona Lasinio. The argument is based on the 
definition of an effective action for the last model and on its comparison 
with the fermion effective action of QED at fixed energy density. I have 
shown how the singularity in the QED effective action, which controls the 
critical behaviour of this model, is not accessible to the Nambu-Jona Lasinio 
model and therefore both models are most likely in different universality 
classes.

The previous expectations are corroborated by the results of the fits of 
the longitudinal susceptibility at the critical point of QED reported 
in section 3. The value of the critical index $\delta= 1.98(16)$ extracted 
from these fits is quite far from its mean field value $\delta= 3$. 
The most reliable scenario is therefore that in which strongly 
coupled QED has a non-Gaussian fixed point and, provided that 
hyperscaling is verified, a non-trivial continuum limit.

\vskip 1truecm
\leftline{\bf Acknowledgements}
\par
Much of the work reported here has been done through a CICYT (Spain)-INFN 
(Italy) collaboration. I wish to thank G. Di Carlo, A. Galante, A.F.~Grillo, 
V. Laliena and C.E. Piedrafita for their invaluable co-operation. I would 
like to thank also A. Koci\'c and J.B. Kogut for very 
interesting discussions.

\vfill
\eject
\centerline{\bf References}
\vskip 1 truecm
\noindent
1. J. Bartholomew, S.H. Shenker, J. Sloan, J.B. Kogut, M. Stone, 
H.W.~Wyld, J. Shigmetsu and D.K. Sinclair, 
Nucl. Phys. {\bf B230} \rm (1984) 222. 
\vskip 0.5truecm
\noindent
2. V.A. Miransky, 
Nuovo Cim. {\bf 90A} \rm (1985) 149.
\vskip 0.5truecm
\noindent
3. C.N. Leung, S.T. Love and W.A. Bardeen, 
Nucl. Phys. {\bf B273} \rm (1986) 649.
\vskip 0.5truecm
\noindent
4. V. Azcoiti, A. Cruz, E. Dagotto, A. Moreo and A. Lugo, 
Phys. Lett. {\bf  B175} \rm (1986) 202.
\vskip 0.5truecm
\noindent
5. J.B. Kogut, E. Dagotto, 
Phys. Rev. Lett. {\bf 59} \rm (1987) 617.
\vskip 0.5truecm
\noindent
6. J.B. Kogut, E. Dagotto and A. Koci\'c, 
Phys. Rev. Lett. {\bf 60} \rm (1988) 772.
\vskip 0.5truecm
\noindent
7. A.M. Horowitz, 
Nucl. Phys. {\bf B17} (Proc. Suppl.) \rm (1990) 694.
\vskip 0.5truecm
\noindent
8. M. G\"ockeler, R. Horsley, P. Rakow, G. Schierholz and R. Sommer, 
Nucl. Phys. {\bf B371} \rm (1992) 713.
\vskip 0.5truecm
\noindent
9. A. Kocic, J.B. Kogut and K.C. Wang, 
Nucl. Phys. {\bf B398} \rm (1993) 405; 
V.~Azcoiti, G. Di Carlo, A.F.Grillo, 
Int. J. Mod. Phys. {\bf  A8} \rm (1993) 4235. 
\vskip 0.5truecm
\noindent
10. A.M. Horowitz, 
Phys. Rev. {\bf D43} \rm (1991) R2461; 
A. Koci\'c and J.B. Kogut, 
Nucl. Phys. {\bf B422} \rm (1994) 593; 
S. Kim, A. Koci\'c and J.B. Kogut,  
Nucl. Phys. {\bf B429} \rm (1994) 407.
\vskip 0.5truecm
\noindent
11. V. Azcoiti, G. Di Carlo, A. Galante, A.F. Grillo, V. Laliena, 
C.E.~Piedrafita, 
Phys. Lett. {\bf B355} \rm (1995) 270.
\vskip 0.5truecm
\noindent
12. V. Azcoiti et al., "Continuous Chiral Transition in Strongly Coupled 
Compact QED with the Standard Torus Topology", to appear.
\vskip 0.5truecm
\noindent
13. V. Azcoiti, G. Di Carlo, A. Galante, A.F. Grillo, V. Laliena 
and C.E.~Piedrafita, 
Phys. Lett. {\bf B379} \rm (1996) 179.
\vskip 0.5truecm
\noindent
14. M. G\"ockeler, R. Horsley, V. Linke, P. Rakow, G. Schierholz and 
H. Stuben, 
\rm hep-lat 9605035 (1996).
\vskip 0.5truecm
\noindent
15. J. Jersak, C.B. Lang and T. Neuhaus, 
\rm hep-lat 9606010 (1996); 
M. Baig and H.~Fort, 
Phys. Lett. {\bf B332} \rm (1994) 428.
\vskip 0.5truecm
\noindent
16. W. Kerler, C. Rebbi and A. Weber, 
\rm hep-lat 9607009 (1996).
\vskip 0.5truecm
\noindent
17. V. Azcoiti, G. Di Carlo, A. Galante, A.F. Grillo, V. Laliena 
and C.E. Piedrafita, 
Phys. Lett. {\bf B353} \rm (1995) 279.
\vskip 0.5truecm
\noindent
18. V. Azcoiti, G. Di Carlo and A.F. Grillo, 
Phys. Rev. Lett. {\bf 65} \rm (1990) 2239;
V. Azcoiti, A. Cruz, G. Di Carlo, A.F. Grillo and A. Vladikas, 
Phys. Rev. {\bf D43} \rm (1991) 3487; 
V. Azcoiti, G. Di Carlo, L.A. Fernandez, A. Galante, 
A.F. Grillo, V. Laliena, 
X.Q. Luo, C.E. Piedrafita and A. Vladikas, 
Phys. Rev. {\bf D48} \rm (1993) 402. 
\vskip 0.5truecm
\noindent
19. E. Dagotto, A. Kocic and J.B. Kogut, 
Phys. Lett. {\bf B231} \rm (1989) 235.
\vskip 0.5truecm
\noindent
20. V. Azcoiti, G. Di Carlo and A.F. Grillo, 
Phys. Lett. {\bf B305} \rm (1993) 275.
\vskip 0.5truecm
\noindent
21. V. Azcoiti, V. Laliena and X.Q. Luo, 
Phys. Lett. {\bf B354} \rm (1995) 111.
\vskip 0.5truecm
\noindent
22. S. Hands, A. Kocic and J.B. Kogut, 
Ann. Phys. {\bf 224} \rm (1993) 29; 
J.A.~Gracey, 
Phys. Lett. {\bf B308} \rm (1993) 65; 
Phys. Rev. {\bf D50} \rm (1994) 2840.
\vskip 0.5truecm
\noindent
23. M.P. Lombardo, A. Koci\'c and J.B. Kogut, 
\rm hep-lat 9411051 (1994).

%#
\vfill
\eject
\vskip 1 truecm
\leftline{\bf Figure captions}
\vskip 1 truecm
{\bf Figure 1.} $ 0.97{\langle\,\bar\psi\psi\,\rangle^3}{\log ({1\over{
\langle\,\bar\psi\psi\,\rangle}})}
+ 2.12{\langle\,\bar\psi\psi\,\rangle^3}$ against $m$ in the $14^4$ 
lattice. The solid line 
is a fit with a straight line of slope equal 
to 1 and crossing the origin.

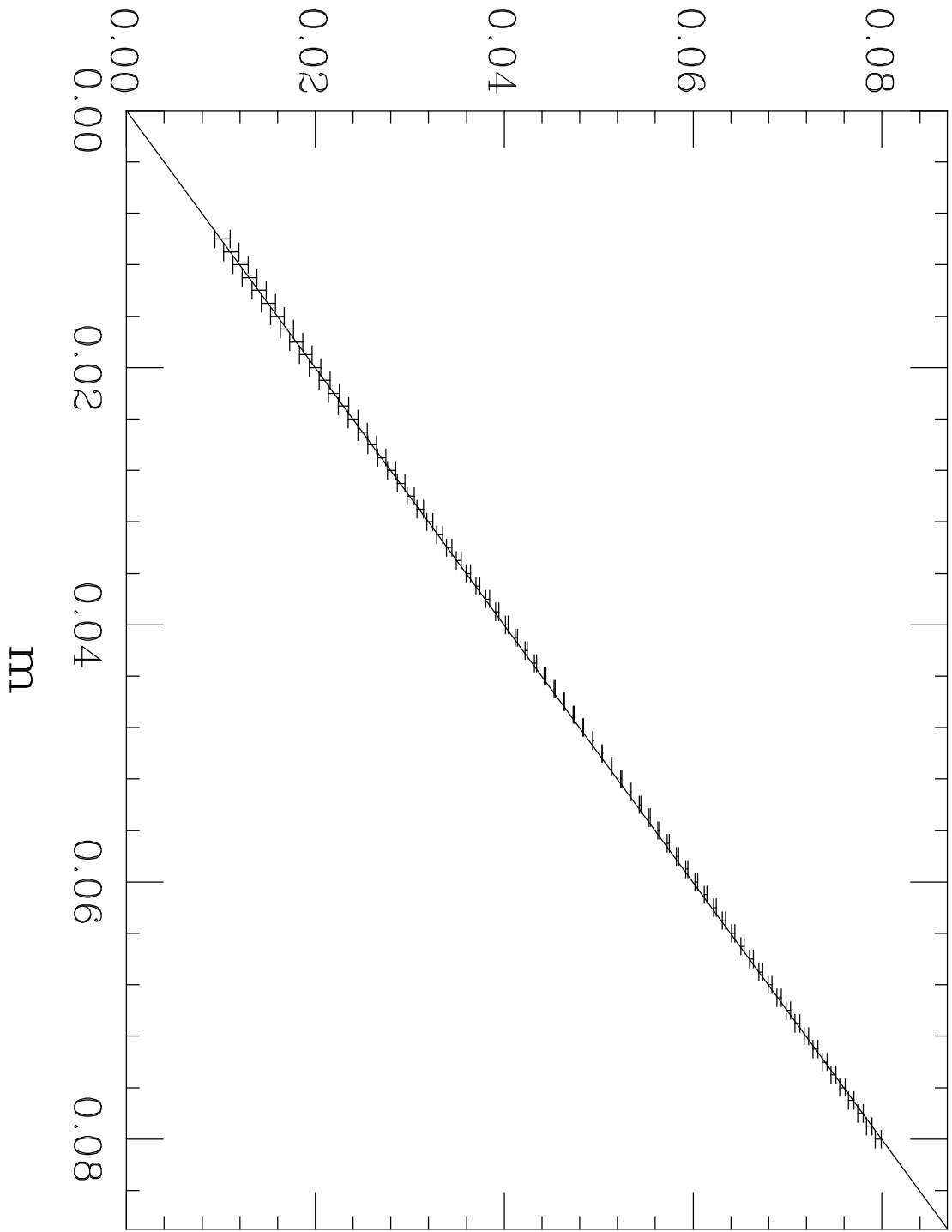
{\bf Figure 2.} $2.39{\langle\,\bar\psi\psi\,\rangle^{2.73}}$
against $m$ in the $14^4$ 
lattice. The solid line 
is a fit with a straight line of slope equal 
to 1 and crossing the origin.

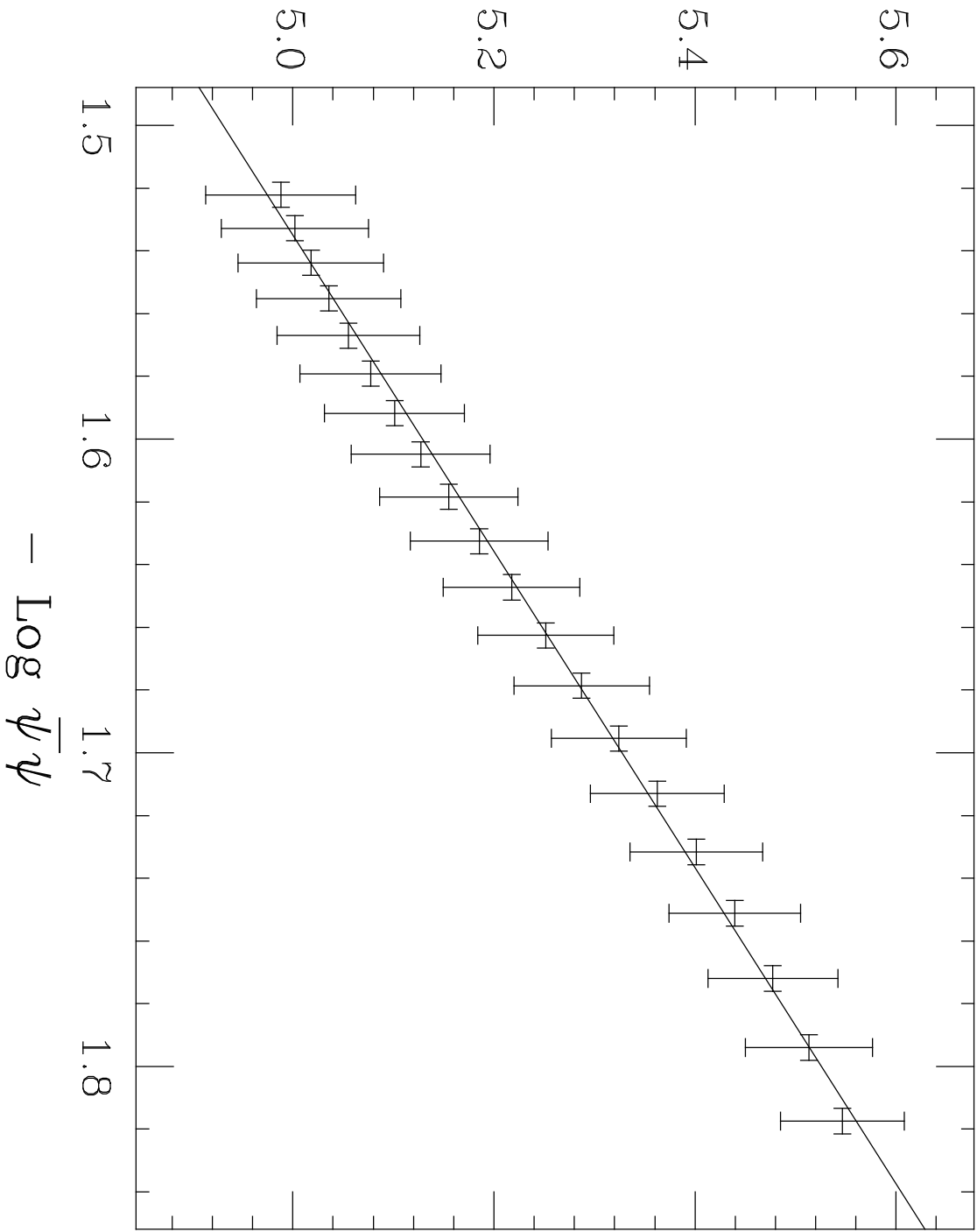
{\bf Figure 3.} $\log {(3m - \langle\,\bar\psi\psi\,\rangle\chi^{-1})}$ 
against $\log {\langle\,\bar\psi\psi\,\rangle}$ at the critical point 
of the $12^4$ lattice.
The solid line is a linear fit.

\vfill
\eject
\vskip 1 truecm
\leftline{\bf Table captions}
\vskip 1 truecm
{\bf Table I} Critical couplings 
extracted from the susceptibilities in $10^4,12^4$ and $14^4$ lattices 
at $N_f=1,2,3,4,8$.

{\bf Table II} Results for the constant $B$ 
and the critical index $\delta$ (pure power law equation (25)) 
and for the two constants $A$ and $C$ (mean field equation (24)) 
at lattice sizes $10, 12, 14$ and $N_f=1,2,3,4,6,8$. All the fits were 
done in the mass interval $(0.01--0.08)$.

\end{document}